\documentclass[journal]{IEEEtran}
\usepackage{graphicx}
\usepackage{bm}
\usepackage{amsmath}
\usepackage{amsfonts}
\usepackage{amsthm}
\usepackage{amssymb}
\usepackage{color}
\usepackage{mathrsfs}
\usepackage[keeplastbox]{flushend}
\usepackage[utf8]{inputenc}
\usepackage{tikz}
\usetikzlibrary{arrows, calc, decorations.markings, positioning}
\usepackage{cite}
\usepackage{booktabs}
\usepackage{multirow}
\usepackage{amsmath}
\usepackage{amsthm}
\usepackage{algorithm}
\usepackage{algorithmic}
\usepackage[flushleft]{threeparttable}
\usepackage{midfloat}
\usepackage{tabularx}
\usepackage{threeparttable}
\ifCLASSOPTIONcompsoc
 \usepackage[caption=false,font=normalsize,labelfont=sf,textfont=sf]{subfig}
\else
 \usepackage[caption=false,font=footnotesize]{subfig}
\fi

\begin{document}
\title{Seventy Years of Radar and Communications: \\The Road from Separation to Integration
}
\author{
	{
	Fan Liu,~\IEEEmembership{Member,~IEEE}, Le Zheng,~\IEEEmembership{Senior Member,~IEEE}, Yuanhao Cui,~\IEEEmembership{Member,~IEEE}, \\ Christos Masouros,~\IEEEmembership{Senior~Member,~IEEE}, Athina P. Petropulu,~\IEEEmembership{Fellow,~IEEE}, \\Hugh Griffiths,~\IEEEmembership{Fellow,~IEEE}, and Yonina C. Eldar,~\IEEEmembership{Fellow,~IEEE}
	} 
\thanks{F. Liu and Y. Cui are with the Southern University of Science and Technology, Shenzhen, China. Y. Cui was with the Beijing University of Posts and Telecommunications, Beijing, China (e-mail: \{cuiyh,liuf6\}@sustech.edu.cn).}
\thanks{L. Zheng is with the Beijing Institute of
Technology (BIT), Beijing, China (e-mail: le.zheng.cn@gmail.com).}
\thanks{C. Masouros and H. Griffiths are with the University College London, London, WC1E 7JE, UK (e-mail: chris.masouros@ieee.org, h.griffiths@ucl.ac.uk).}
\thanks{A. P. Petropulu is with the Rutgers, the State University of New Jersey, NJ 08854, United States (e-mail: athinap@rutgers.edu).}
\thanks{Y. C. Eldar is with the Weizmann Institute of Science, Rehovot, Israel (e-mail: yonina.eldar@weizmann.ac.il).}
}
\maketitle

\begin{abstract}
Radar and communications (R\&C) as key utilities of electromagnetic (EM) waves have fundamentally shaped human society and triggered the modern information age. Although R\&C have been historically progressing separately, in recent decades they have been converging towards integration, forming integrated sensing and communication (ISAC) systems, giving rise to new, highly desirable capabilities in next-generation wireless networks and future radars. To better understand the essence of ISAC, this paper provides a systematic overview on the historical development of R\&C from a signal processing (SP) perspective. We first interpret the duality between R\&C as signals and systems, followed by an introduction of their fundamental principles. We then elaborate on the two main trends in their technological evolution, namely, the increase of frequencies and bandwidths, and the expansion of antenna arrays. We then show how the intertwined narratives of R\&C evolved into ISAC, and discuss the resultant SP framework. Finally, we overview future research directions in this field.
\end{abstract}

\section{Introduction}
\subsection{Background and Motivation}
\IEEEPARstart {S}{ince} the 20th century, the development of human civilization has relied largely upon the exploitation of electromagnetic (EM) waves. Governed by Maxwell's equations, EM waves are capable of travelling over large distances at the speed of light, which makes them a perfect information carrier. In general, one may leverage EM waves to acquire information on physical targets, including range, velocity, and angle, or to efficiently deliver artificial information, e.g., texts, voices, images, and videos from one point to another. Among many applications, EM waves have enabled information acquisition and delivery, which form the foundation of our modern information era, and have given rise to the proliferation of radar and communication (R\&C) technologies.

While the existence of EM waves was theoretically predicted by Maxwell in 1865, and experimentally verified by Hertz in 1887, its capability of carrying information to travel long distances was not validated until Marconi's transatlantic wireless experiment in 1901 \cite{griffiths2018oliver}. The successful reception of the first transatlantic radio signal marked the beginning of the great information era. From then on, communication technology has rapidly grown thanks to the heavy demand for intelligence, intercept and cryptography technologies during the two world wars. It is generally difficult to identify a precise date for the birth of radar. Some of the early records showed that the German inventor C. Hülsmeyer was able to use radio signals to detect distant metallic objects as early as 1904. In 1915, the British radar pioneer Robert Watson Watt, employed radio signals to detect thunderstorms and lightning. The R\&D of modern radar systems was not carried out until the mid 1930s. The term RADAR was first used by the US Navy as an acronym of ``RAdio Detection And Ranging" in 1939.

Despite the fact that both technologies originated from the discoveries of Maxwell and Hertz, R\&C have been largely treated as two separate research fields due to different constraints in their respective applications, and were therefore independently investigated and developed for decades. Historically, the technological evolution of R\&C follows two main trends: a) from low frequencies to higher frequencies and larger bandwidths \cite{1359140}, and  b) from single-antenna to multi-antenna or even massive-antenna arrays \cite{4350230,5595728}. With recent developments, the combined use of large-antenna arrays and Millimeter Wave (mmWave)/Terahertz (THz) band signals results in striking similarities between R\&C systems in terms of the hardware architecture, channel characteristics, as well as signal processing methods. Hence, the boundary between R\&C is becoming blurred, and hardware and spectrum convergence has led to a design paradigm shift, where the two systems can be co-designed for efficiently utilizing resources, offering tunable tradeoffs and unprecedented synergies for mutual benefits. 
This line of research is typically referred to as integrated sensing and communications (ISAC), and is applicable in numerous emerging areas, including vehicular networks, IoT networks, and activity recognition \cite{9737357,8999605}. Over the last decade, ISAC has been well-recognized as a key enabling technology for both next-generation wireless networks and radar systems \cite{9737357}. Given the potential of ISAC, a deeper understanding of the various connections and distinctions between R\&C, and learning from how they evolved from separation to integration, is important for inspiring future research.

\begin{figure*}
    \centering
    \includegraphics[width=2\columnwidth]{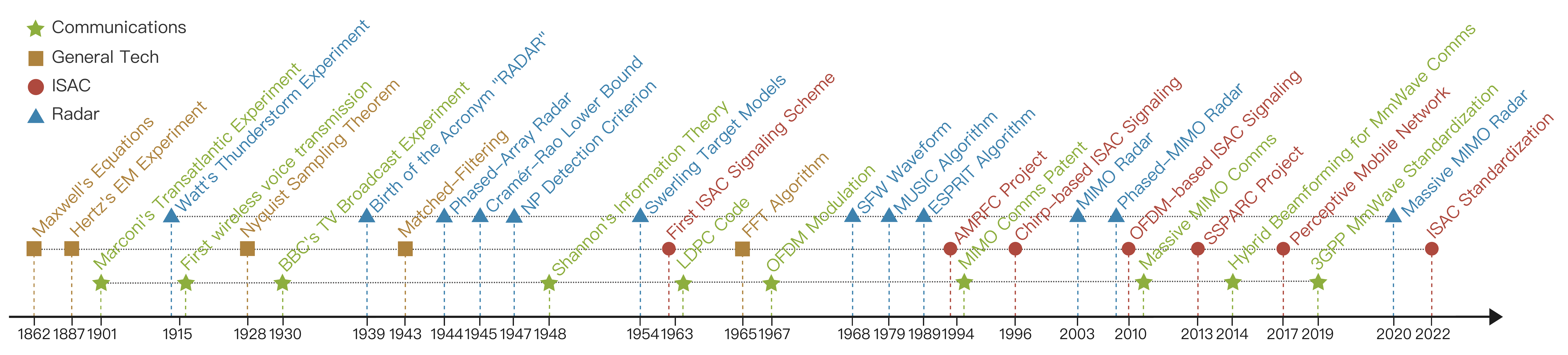}
    \caption{Important milestones for radar and communications signal processing.}
    \label{fig: RC Milestones}
\end{figure*}

In Fig. \ref{fig: RC Milestones} we summarize key milestones achieved in R\&C history, which are split into four categories with different markers, namely, individual R\&C technologies, general technologies that are useful for both, and ISAC technologies. In the remainder of the paper, we will discuss how these key techniques facilitate the development of R\&C and ISAC systems.

\subsection{Summary and Organization of the Paper}
In this paper, we provide a systematic overview on the development and key milestones achieved in the history of R\&C from an SP perspective. We commence by introducing the fundamental principles and SP theories of both R\&C. We then present the spectrum engineering of R\&C, namely, from narrowband to wideband, and from single-carrier to multi-carrier systems. Furthermore, we elaborate on the expansion of R\&C systems' antenna arrays, i.e., from single-antenna systems, to phased-array, and to MIMO, massive MIMO (mMIMO), and distributed antenna systems. Following the above two technological trends, the paths of R\&C eventually move from separation to integration, and give rise to the ISAC technology, which motivates the detailed discussion on the SP framework of ISAC. Finally, we summarize the paper and identify future research directions.



\section{Fundamentals of Radar and Communications}
\subsection{Basic Principles: A Signals-and-Systems Perspective}

The basic system setting for both R\&C consists of three parts: a transmitter (Tx), which produces EM waves, a channel, over which EM waves propagate, and a receiver (Rx), which receives EM waves distorted by the channel. While communication Txs and Rxs are usually well separated, radar Txs and Rxs may either be collocated or separately positioned, leading to mono-static or bi-static radar settings, respectively. In more complicated scenarios, multiple Txs and Rxs may be involved in both applications, which correspond to multi-user communications and multi-static radar systems.

It is often convenient to represent EM waves by the electrical field intensity, as a complex signal as a function of time $t$. The core tasks for R\&C can then be defined as:
\begin{itemize}
    \item {\textbf{Information Acquisition for Radar:}} The aim here is to extract the target information embedded in the received signal, given knowledge of the transmit signal.
    \item {\textbf{Information Delivery for Communications:}} The aim here is to recover the useful information contained in the transmit signal at the communication Rx, with knowledge of the channel response.
\end{itemize}

By denoting the signals at the Tx and Rx at time $t$ as $s\left(t\right)$ and $y\left(t\right)$, respectively, the propagation of the signal within the channel can be modeled as a mapping from its input $s\left(t\right)$ to the output $y\left(t\right)$. Ideally, if the noise and disturbance are not considered, such a mapping is {\textit{linear}} due to the physical nature of EM fields and waves, or equivalently, owing to the linearity of Maxwell's equations. Furthermore, if the channel characteristics remain unchanged within a certain time period, it can be approximated as a linear time-invariant (LTI) system, characterized by its impulse response $h\left(t\right)$. As such, the linear mapping is expressed as a convolution integral $y\left( t \right) = \left(s*h\right)\left( t \right)$. While the signaling pulses may be of different forms for R\&C, we suppose that a Nyquist pulse is leveraged such that $s\left(t\right)$ is substantially time-limited on a finite interval $\left[-T,T\right]$. Therefore, signal can be sampled in a nearly lossless manner after passing through the pulse-shaping filter at the Rx, expressed as a convolution sum $y\left( n \right) = \left(s*h\right)\left( n \right)$ at the $n$th sampling point. Let ${\mathbf{s}} = {\left[ {s\left( -N\right), \ldots ,s\left(N\right)} \right]^T}$ be the Tx signal with length $2N + 1$, ${\mathbf{h}} = {\left[ {h\left( 0 \right), \ldots ,h\left( {P - 1} \right)} \right]^T}$ be the channel impulse response with length $P$, and ${\mathbf{y}} = {\left[ {y\left( -N\right), \ldots ,y\left(N+P-1 \right)} \right]^T} $ be the Rx signal with length $2N+P$. Then the convolution can be recast as $\mathbf{y} = \mathbf{H}\mathbf{s}$, where ${\mathbf{H}} = \operatorname{Toep} \left( {\mathbf{h}} \right) \in \mathbb{C}^{\left(2N+P\right) \times \left(2N+1\right)}$ is a Toeplitz matrix, with the $n$th column being $\left[{\mathbf{0}}_{n-1}^T,{\mathbf{h}}^T,{\mathbf{0}}_{2N-n+1}^T\right]^T$.
Alternatively, one may express $\mathbf{y}$ as $\mathbf{y} = \mathbf{S}\mathbf{h}$ by the commutative property, where ${\mathbf{S}} = \operatorname{Toep} \left( {\mathbf{s}} \right) \in \mathbb{C}^{\left(2N+P\right) \times P}$.

The above duality between interchangeable signals and systems implies an interesting connection between R\&C. From the communication perspective, the process of the Tx signal passing through a channel may be viewed as a linear transform $\mathbf{H}$ applied to $\mathbf{s}$, with the communication task being to recover the information embedded in $\mathbf{s}$ by receiving $\mathbf{y}$. From the radar perspective, the sensing task is to recover the target parameters embedded in $\mathbf{h}$, which is viewed as an input ``signal", by observing $\mathbf{y}$, which is viewed as an output signal linearly transformed from $\mathbf{h}$ through a ``system" $\mathbf{S}$. This reveals that the basic SP problems in R\&C are mathematically similar. 

\subsection{Linear Gaussian Models}
Consider the more general linear Gaussian signal model by taking additive white Gaussian noise (AWGN) into account:
\begin{equation}\label{linear_Gaussian}
{\mathbf{Y}} = {\mathbf{H}}\left( {\bm{\eta }} \right){\mathbf{S}}\left( {\bm{\xi }} \right) + {\mathbf{Z}},
\end{equation}
where $\mathbf{Y}$ and $\mathbf{S}$ are the sampled receive and transmit signals, which could be defined over multiple domains, e.g., time-space or time-frequency domain, $\mathbf{H}$ is the corresponding channel matrix (not necessarily Topelitz), and $\mathbf{Z}$ is the white Gaussian noise signal with variance $\sigma^2$. 
The channel $\mathbf{H}$ is a function of the physical parameters $\bm{\eta}$, e.g., range, angle, and Doppler. The transmit signal $\mathbf{S}$ may be encoded/modulated with some information codewords ${\bm \xi}$. Model (\ref{linear_Gaussian}) represents many R\&C systems as elaborated below.
\begin{itemize}
    \item {\textbf{Radar Signal Model:}} Radar systems aim at extracting target parameters $\bm \eta$ from $\mathbf{Y}$. For both radar Tx and Rx, $\mathbf{S}$ is typically a known deterministic signal, in which case $\bm \xi$ can be omitted since the radar waveform contains no information. This can be expressed as
    \begin{equation}\label{radar_general_model}
    {{\mathbf{Y}}_{\rm r}} = {{\mathbf{H}}_{\rm r}}\left( {\bm{\eta }} \right){{\mathbf{S}}_{\rm r}} + {{\mathbf{Z}}_{\rm r}}. 
    \end{equation}
    \item {\textbf{Communication Signal Model:}} Communication systems aim at recovering codewords $\bm\xi$ from $\mathbf{Y}$. The channel $\mathbf{H}$, which is sometimes regarded as an unstructured matrix, can be estimated {\textit{a priori}} via pilots. Therefore, knowing $\bm \eta$ may not be the first priority. 
The resulting model is
    \begin{equation}\label{comms_general_model}
     {{\mathbf{Y}}_{\rm c}} = {{\mathbf{H}}_{\rm c}}{{\mathbf{S}}_{\rm c}}\left( {\bm\xi}  \right) + {{\mathbf{Z}}_{\rm c}}. 
    \end{equation}
\end{itemize}
The subscripts $\left(\cdot\right)_{\rm r}$ and $\left(\cdot\right)_{\rm c}$ are  to differentiate R\&C signals, channels, and noises, respectively. We highlight that (\ref{radar_general_model}) and (\ref{comms_general_model}) describe a variety of R\&C signal models. For example, (\ref{radar_general_model}) can be viewed as the target return of a multi-input multi-output (MIMO) radar in a given range-Doppler bin, where $\bm \eta$ represents angles of targets. Similarly, (\ref{comms_general_model}) may be considered as a narrowband MIMO communication signal. Alternatively, both (\ref{radar_general_model}) and (\ref{comms_general_model}) can be viewed as orthogonal frequency-division multiplexing (OFDM) signal models for R\&C, respectively. In the following, we do not specify the signal domain but focus on generic models (\ref{radar_general_model}) and (\ref{comms_general_model}). More concrete signal models will be discussed in Secs. III-IV.
In addition to individual R\&C systems, (\ref{linear_Gaussian}) may also characterize the general ISAC signal model. That is, a unified ISAC signal serves for dual purposes of information delivering and target sensing, whereas R\&C channels may differ from each other. More details on ISAC systems will be discussed in Sec. V.

\subsection{Fundamental Signal Processing Theories}
Below we elaborate on the fundamental SP theories of R\&C, and in particular focus on (\ref{radar_general_model}) and (\ref{comms_general_model}). 
\subsubsection{Signal Detection}
Signal detection problems arise from many R\&C applications. One essential task for radar is to determine whether a target exists by observing $\mathbf{Y}_{\rm r}$, modeled as a binary hypothesis testing (BHT) problem
\begin{equation}\label{BHT}
    {\mathbf{Y}_{\rm r}} = \left\{ \begin{gathered}
  {\mathcal{H}_0}:{\mathbf{Y}_{\rm r}} = {\mathbf{Z}_{\rm r}}, \hfill \\
  {\mathcal{H}_1}:{\mathbf{Y}_{\rm r}} = {\mathbf{H}_{\rm r}}\left( {\bm\eta}  \right){\mathbf{S}_{\rm r}} + {\mathbf{Z}_{\rm r}}, \hfill \\ 
\end{gathered}  \right.
\end{equation}
where $\mathcal{H}_0$ represents the null hypothesis, i.e., the radar receives nothing but noise, and $\mathcal{H}_1$ stands for the hypothesis where the radar receives both the target return and noise. To address the BHT problem above, one may need to design a {\it{detector}} $\mathcal{T}\left(\cdot\right)$ that maps the received signal $\mathbf{Y}_{\rm r}$ to a real number, and then compare the output with a preset threshold $\gamma$, to determine which hypothesis to choose as true. A target detector may, for example, maximize the detection probability $P_D = \text{Pr}\left({\mathcal{H}_1}\left| {\mathcal{H}_1} \right.\right)$, while maintaining a low false-alarm probability $P_{FA} = \text{Pr}\left({\mathcal{H}_1}\left| {\mathcal{H}_0} \right.\right)$, following the Neyman-Pearson (NP) criterion \cite{kotz2012breakthroughs}.

Signal detection also plays a critical role at the communication Rx. In (\ref{comms_general_model}), the communication Rx observes ${{\mathbf{Y}}_{\rm c}} = {{\mathbf{H}}_{\rm c}}{{\mathbf{S}}_{\rm c}}\left({\bm \xi}\right) + {\mathbf{Z}}_{\rm c}$, and seeks to yield an estimate $\hat{\bm\xi}$ of the information symbol vector ${\bm\xi}  = {\left[ {{\xi _1},{\xi _2}, \ldots ,{\xi _N}} \right]^T} \in \mathcal{A}$. This problem can be solved by leveraging the minimum error probability (MEP) criterion. That is, to minimize the error probability ${P_e} = \sum\nolimits_{i = 1}^{\left| {{\mathcal{A}}} \right|} {\Pr \left( {{{\hat \xi }_i} \ne {\xi _i}} \right)\Pr \left( {{\xi _i}} \right)}$, where $\left| {{\mathcal{A}}} \right|$ is the cardinality of $\mathcal{A}$. The MEP criterion can be translated to the MAP criterion, i.e., the recovered symbols should be the maximizer of the {\it {a posterior}} probability. Note that the decision region in the MEP criterion for communication symbols is determined by their {\it {a priori}} probability, while the decision thresholds in the NP criterion for radar is determined by the required false-alarm probability, resulting in different designs for R\&C detectors.
\subsubsection{Parameter Estimation}
Parameter estimation represents another category of basic SP techniques in R\&C systems. For a radar system, once a target is confirmed to be present, it needs to further extract its parameters ${\bm \eta}$ from $\mathbf{Y}_{\rm r}$ by conceiving an {\it{estimator}}, mapping $\mathbf{Y}_{\rm r}$ from the signal space to an estimate ${\hat{\bm \eta}}$, defined as ${\hat {\bm\eta}}  = \mathcal{F}\left( {{{\mathbf{Y}}_{\rm r}}} \right)$.
To measure how accurate an estimator is, a possible performance metric is the mean squared error (MSE), expressed as $\varepsilon  = \mathbb{E}\left( {{{\left\| {{\mathbf{\bm \eta }} - {\mathbf{\hat {\bm \eta} }}} \right\|}^2}} \right)$. The average may be over the noise or also over the parameters if they are assumed random. When the parameters are assumed to be deterministic, the MSE of any unbiased estimate is lower-bounded by the Cram\'er-Rao bound (CRB), defined as the inverse of the Fisher information matrix $\mathbf{J}$ \cite{kotz2012breakthroughs}
\begin{equation}
    \mathbb{E}{\left[ {\left( {{\bm{\eta }} - {\bm{\hat \eta }}} \right)\left( {{\bm{\eta }} - {\bm{\hat \eta }}} \right)^H} \right]} \succeq {{\mathbf{J}}^{ - 1}} =  \left\{- \mathbb{E}\left[{\frac{{{\partial ^2}\ln p\left( {{{\mathbf{Y}}_{\rm r}};{\bm{\eta }}} \right)}}{{\partial {{\bm{\eta }}^2}}}} \right]\right\}^{-1},
\end{equation}
where $p\left( {{{\mathbf{Y}}_{\rm r}};{\bm{\eta }}} \right)$ is the probability density function (PDF) of $\mathbf{Y}_{\rm r}$ parameterized by $\bm \eta$. While the maximum likelihood estimate (MLE) asymptotically achieves the CRB, attaining the MLE can be highly computationally expensive. To that end, low-complexity parameter estimation algorithms, e.g., MUSIC and ESPRIT \cite{correct_MUSIC1,32276}, have been widely applied in practical situations, such as angle of arrival estimation.

In communication systems, the channel $\mathbf{H}_{\rm{c}}$ should be estimated before delivering the useful information. For channel estimation, the Tx sends pilots to the Rx, which are reference signals known to both. The Rx then estimates the channel based on both received signals and pilots. Channel estimation is mathematically similar to the target estimation problem, where the to-be-estimated parameters $\bm \eta$ are entries of $\mathbf{H}_{\rm c}$, which is regarded as an unstructured matrix. We will elaborate on similarities and differences between estimation tasks for communication channels and radar targets in Sec. IV.

\subsubsection{Information Theory}
Information theory serves as the foundation of communication SP. A remarkable result attained by C. E. Shannon in his landmark paper \cite{shannon1948mathematical}, published in 1948, states that, for any discrete memoryless channel (DMC) with input $X$ and output $Y$, the {\it{channel capacity}} is $C = \mathop {\max }\limits_{p\left( X \right)} \;I\left( {X;Y} \right)$, where the maximum is taken over all possible input distribution $p\left(X\right)$, and $I\left( {X;Y} \right)$ is the mutual information (MI) between $X$ and $Y$. The channel coding theorem states that a coding rate $R$ below $C$ is achievable. Conversely, if $R > C$, arbitrarily small decoding error is not possible. Information theory may also be adopted to measure the radar performance \cite{1188560}, and may reveal profound connections between R\&C. Let us consider a generic real-valued Gaussian channel with an input $X$, which is assumed to be random, and output $Y$. In the communication case, $X$ can be an information-carrying signal emitted by the Tx, and $Y$ the signal received at the communication Rx. In the radar case, $X$ can be some random target parameter/channel to be estimated, and $Y$ the echo signal received at the radar Rx. In both R\&C tasks, we may wish to accurately/approximately recover $X$ by observing $Y$.

We denote the mutual information between $X$ and $Y$ as $I\left(X;Y\right)$, and the minimum MSE (MMSE) of estimating $X$ from $Y$ as  $\operatorname{MMSE}\left(X|Y\right) = \mathbb{E}\left\{|X - \mathbb{E}\left(X|Y\right)|^2\right\}$, both of which may be expressed as functions of the SNR, namely, $I\left( {\operatorname{snr} } \right)$ and $\operatorname{MMSE} \left( {\operatorname{snr} } \right)$. We then have the following I-MMSE identity holds for Gaussian channels\cite{1412024}
\begin{equation}\label{I-MMSE}
\frac{d}{{d\operatorname{snr} }}I\left( {\operatorname{snr} } \right) = \frac{1}{2}\operatorname{MMSE} \left( {\operatorname{snr} } \right).
\end{equation}
The above relationship implies that, the increasing rate (derivative) of the mutual information between $X$ and $Y$ with respect to the SNR is half of the MMSE for estimating $X$ given $Y$. For a Gaussian channel, $I\left( {\operatorname{snr} } \right)$ is maximized by inputting a Gaussian distributed $X$ under a given SNR. More precisely, a Gaussian input always results in the most rapidly growing mutual information, and, accordingly, yields the maximum MMSE, making it the most favorable for communication yet the least favorable for radar sensing. From a communication perspective, the channel input should be ``as random as possible" to carry more information. From a radar perspective, estimation performance becomes more inaccurate if the target parameters change more randomly. The Gaussian distribution has the highest entropy (randomness) under a second-order moment constraint (i.e., fixed power budget), resulting in this interesting tradeoff.

\subsection{Interplay between R\&C}
While communication happens between cooperative Txs and Rxs, radar sensing is essentially uncooperative, even if the radar Tx and Rx are colocated. This distinction results in inherently different R\&C SP frameworks. First, both R\&C signal processing aims at recovering the useful information contained in the received signal with minimum distortion. The communication system, however, needs another level of performance guarantee, i.e., to transmit, receive, and actively control as much information as possible. This requires sophisticatedly tailored encoding and decoding, and modulation and demodulation strategies at the Tx and Rx, respectively, which motivates the development of information theory, whose spirit forms the foundation of modern communication SP framework. Moreover, as the communication Tx and Rx are highly cooperative, they are able to share the SP complexities in a rather flexible manner, depending on the specific scenarios. For instance, in a downlink communication setup where powerful base station (BS) sends information to the user, most of the complicated signal processing is done at the Tx's side, e.g., precoding, thus to ease the computational burden at the user's side. In a radar system, however, the complexity of the Rx SP always dominates its Tx counterpart, yet they are typically unable to share design complexities.

In what follows, we elaborate on the evolution of R\&C in terms of both the spectrum engineering and antenna array technologies, and further reveal their interplay in the spectral and spatial SP.

\section{Spectrum Engineering: The Road to Higher Frequency and Larger Bandwidth} 
\subsection{Spectrum Characteristics and Management}
\label{subsec:spectrumcharac}

\begin{figure*}
    \centering
    \includegraphics[width=2\columnwidth]{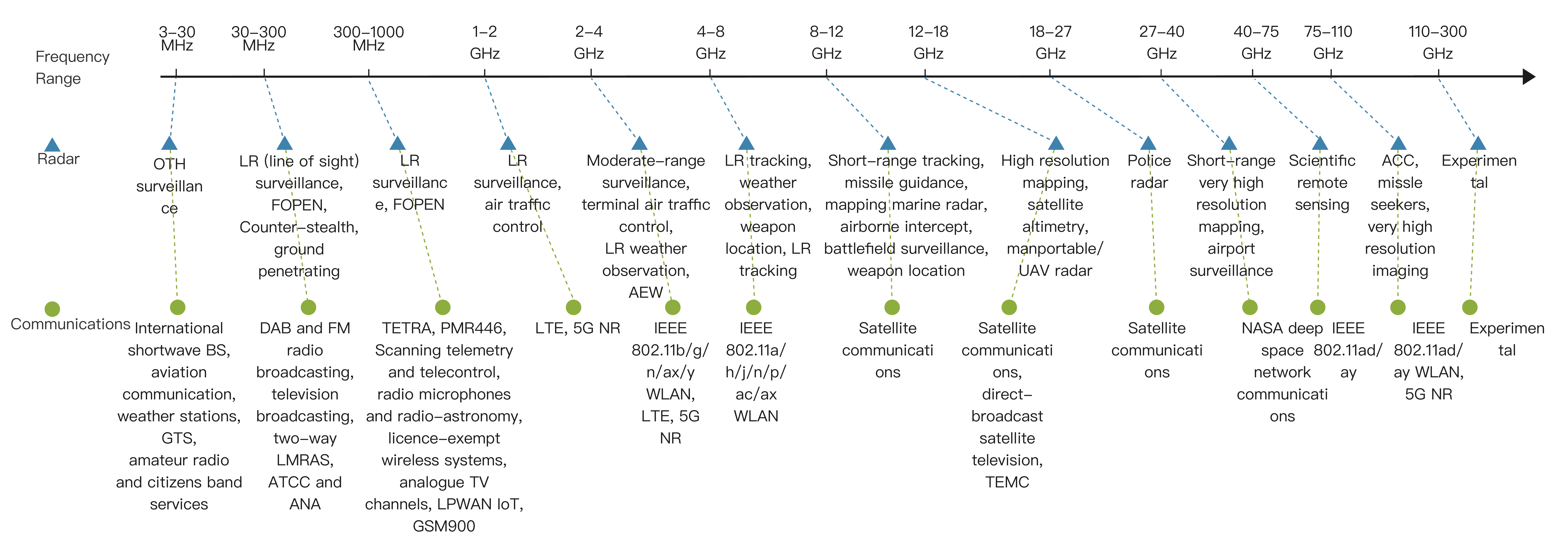}
    \caption{Summary of frequency bands and their usage in R\&C applications. {\textit{Abbreviations:}} Airborne Early Warning (AEW), Automobile Cruise Control (ACC), Unmanned Aerial Vehicle (UAV), Long-Range (LR), Foliage Penetration (FOPEN), Ground Penetrating Radar (GPR), Over-the-Horizon (OTH), Base Station (BS), Air Traffic Control Communications (ATCC), Land Mobile Radio Systems (LMRAS), Terrestrial Trunked Radio (TETRA), Terrestrial Microwave Communications (TEMC), Government Time Stations (GTS), Air Navigation (ANA). }
    \label{fig: freqbands}
\end{figure*}

The radio-frequency (RF) EM spectrum, extending from below 1 MHz to above 100 GHz, has been used for a wide range of applications, including communications, radio and television broadcasting, radio-navigation, and sensing \cite{6967722_1}. Fig. \ref{fig: freqbands} lists the frequency bands where R\&C systems operate and highlights the modes and usage that are performed in each band. For radar sensing, the lower bands offer some unique capabilities such as long-range surveillance and weather monitoring\cite{6967722_1}. For communications, lower bands exhibit low signal attenuation, making them suitable for long-distance transmission. 

The higher frequency bands provide some advantages to R\&C. For a fixed fractional bandwidth, increasing the operating frequency subsequently increases the achievable bandwidth, thus providing finer range resolution for radar and higher data rates for communications. However, in these higher bands, long-range operation becomes more strongly affected by attenuation due to the atmosphere. Moreover, the diffraction effect of high-frequency EM wave signals decreases, which leads to a reduction in the number of paths propagated. As such, radar sensing and wireless communication via these bands are limited to short-range applications. For example, radars from X- to W-bands are used for automotive collision avoidance, police radar, airport surveillance, and scientific remote sensing. As for the communication, the mmWave band are soon to be finalized as part of the 5G NR standards and has been exploited by the 802.11ad/ay WLAN protocols. More advanced radar SP tasks, such as real-time range-Doppler imaging and target recognition, typically rely on sparse recovery methods, as sparse channels are usually required in radar applications. For communication with high frequency or wideband, algorithms are required to be specifically conceived for channel estimation and demodulation to achieve higher data rates.

As a representative wideband signaling strategy, multi-carrier technologies have been extensively applied in both R\&C systems, which we overview in the following.

\subsection{Signal Models and Processing Techniques}
\subsubsection{Multi-Carrier Radar Signal Processing}
Let us consider a pulsed radar, with a non-zero support $[0, \tau]$ for each pulse. The radar works by transmitting a short burst of energy, or pulse, towards the target and then listening for the echo that bounces back. The pulse repetition interval (PRI) is $T_{\text{PRI}}$ and the total transmit bandwidth available at the baseband is $B_{\rm r}$, resulting in a duty cycle of $\tau/T_{\text{PRI}}$. The carrier frequency $f_n$ of the $n$th pulse is chosen from $[f_c - \Delta B/2, f_c + B_{\rm r}- \Delta B/2]$ for the multi-carrier radar system, where $f_c$ is the lowest carrier frequency within the band, and $\Delta B$ is the bandwidth of each subpulse. Specifically, for single-carrier systems, we have $f_n = f_c$ for all $n$ with $f_c \gg B_{\rm r}$. The $n$-th transmit pulse is 
\begin{equation} s_{{\rm r},n}(t) = \sqrt{P_{\rm r}} x_{\rm r}(t-n T_{\text{PRI}} )e^{j 2 \pi f_n (t-n T_{\text{PRI}})},
\end{equation} 
where $P_{\rm r}$ is the radar transmit power. For the linear frequency modulated (LFM) waveform, we have $x_{\rm r}(t) = e^{j \pi B_{\rm{r}} t^2/\tau} {\rm rect}\left( t/\tau \right)$, where ${\rm rect}\left( t/\tau \right)$ is 1 for $0 \leq t \leq \tau$, and 0 elsewhere. The target response is 
\begin{equation}
\label{eq:ht}
h(t) = \sum_{l=0}^{L-1} \alpha_l e^{j 2 \pi \nu_l t} {\bar\delta}(t - \tau_l),
\end{equation}
where ${\bar\delta}(\cdot)$ is the Dirac delta function, $\alpha_l$ is the reflection coefficient, $\tau_l$ and $\nu_l$ are the delay and Doppler of the $l$-th target corresponding to its range and velocity. The time delay between the transmitted and received signal is used to calculate the distance to the target. In general, the radar cannot separate the two targets in range if $|\tau_{l_1} - \tau_{l_2}| < 1/B_{\rm r}$. In many sensing problems, obtaining information at high range resolution is crucial to distinguish closely spaced targets \cite{levanon2004radar}, which incurs larger bandwidth needs.

In 1968, K. Ruttenburg and L. Ghanzi proposed the stepped frequency waveform (SFW) that can be viewed as a form of inter-pulse phase coding \cite{ruttenberg1968high}. It transmits a series of linearly increasing or decreasing frequency signals, or steps, towards the target. The frequency of the received signal is compared to the frequency of the transmitted signal to calculate the distance to the target. By sweeping through a range of frequencies, the radar can also measure the target's speed. SFW was later used in sets of radars, in which coherent integration of a burst of pulses yields high range resolution. Conventional SFW sets the carrier frequency sequence as $f_n = f_c + n \Delta f$ for all $n$. To improve the data rate and avoid interference, more recent approaches randomly draw frequencies from the set ${\cal F} = \left\{ f_n | f_n = f_c + d_n \Delta f \right\}$, where $\Delta f$ is a step size, and $d_n \in {\mathbb Z}$ is chosen from a subset of $[0, D]$ so that $D \Delta f > B$ is the synthesized bandwidth. 

Conventional SFW signal processing follows the matched-filtering (MF) process, in which ${\mathbf Y}_{\rm r}$, ${\mathbf S}_{\rm r}$, and ${\mathbf H}_{\rm r}({\bm \eta})$ are the Rx and Tx signals, and target response in the frequency domain, respectively. With this, we may represent the discretized signal as $\mathbf{y} = \mathbf{S}\mathbf{h} + \mathbf{z}$, with ${\mathbf h}$ being the time-domain target response and $\mathbf{S}$ being the Toeplitz matrix composed of the transmitted signal. For sparse SFW, MF may lead to high sidelobes due to the vacancy in frequency bands. To mitigate the effects of sidelobes, radar designers can use a variety of techniques, such as antenna designs that minimize sidelobes, signal processing algorithms that filter out sidelobes, or adjusting the radar's operating parameters to avoid sidelobe interference. More recent approaches consider the sparse nature of radar signals and estimate the target parameters using sparse recovery algorithms. Assuming that the targets are composed of very few scatterers compared with the number of measurements, $\mathbf h$ can be estimated by solving the optimization problem
\begin{equation}
\label{eq:recover}
\mathop {\min }\limits_{\mathbf{h}} \;{\left\| {\mathbf{h}} \right\|_0}\;{\text{s.t.}}\;{\text{ }}{\left\| {{\mathbf{y}} - {\mathbf{Sh}}} \right\|_2} \le \zeta, 
\end{equation}
where $\zeta$ is a positive constant dependent on the noise variance. This problem can be solved using compressed sensing (CS) algorithms, e.g., $\ell_1$-norm minimization, and greedy algorithms such as orthogonal matching pursuit (OMP) \cite{eldar2012compressed,de_maio_eldar_haimovich_2019}. 

\subsubsection{Multi-Carrier Communication Signal Processing}
As for the communication system, we assume it occupies a frequency band of $B_{\rm{c}}$. Setting $T_c = \frac{1}{B_{\rm{c}}}$, the radiated signal is given by 
\begin{equation}\label{single_carrier_model}
s_{\rm c}(t) = \sum_{n = 0}^{{N_s} - 1}  \sqrt{P_{\rm c}} x_{\rm c}(n)\\ \psi_c(t-nT_{\rm c}) e^{j 2 \pi f_c t}, 
\end{equation}
where $P_{\rm c}$ is the transmit power, $x_{\rm c}(n)$ for all $n$ is the symbol sequence to be transmitted with length $N_s$, and $\psi_c(\cdot)$ satisfies the Nyquist criterion with respect to $T_{\rm c}$. Classic amplitude shift keying (ASK), frequency-shift keying (FSK), phase shift keying (PSK) could be applied for generating $x_{\rm c}(n)$. 

The model in (\ref{single_carrier_model}) is a single-carrier system, which has limitation in bandwidth and data rates. Following a 1965 article, Zimmerman and Kirsch designed a high frequency radio multi-carrier transceiver \cite{zimmerman1967gsc}. When the structure in signal space relies on multiple subcarriers, it corresponds to a multi-carrier scheme represented by letting $ x_{\rm c}(n) = \sum_{m=0}^{N_c - 1} x_{{\rm c},m}(n) \psi_{{\rm{c}},m}(t - n T_{\rm c}) $. Here $x_{{\rm c},m}(n)$ is the symbol sequence being transmitted, $N_c$ is the number of subcarriers, and $\psi_{{\rm{c}},m}(t)$ is the synthesis function which satisfies the Nyquist criterion with respect to $\frac{1}{B_{\rm{c}}}$ and maps $x_{{\rm c},m} (n)$ into the signal space. The family of $\psi_{{\rm{c}},m}(t) = \omega_{\rm{c}}(t) e^{j 2 \pi m \Delta f t}$ is referred to as a Gabor system, where $\omega_{\rm{c}}(t)$ is the prototype filter, and $\Delta f$ is the subcarrier spacing. It is easy to show that an $N_c $-point inverse discrete Fourier transform (IDFT) operating on the data generates samples of the OFDM signal, which can be accelerated by the fast Fourier transform (FFT) algorithm proposed by Cooley \textit{et al.} \cite{cooley1965algorithm}. At the communication Rx, we remove the cyclic prefix (CP) and take the signal samples for $n = 0,1,...,N_s-1$, yielding
\begin{equation}
    {\mathbf S} = {\mathbf F}_{N_c}^H \left( {\mathbf X}_{\rm c}  \odot {\mathbf b}(\tau') {\mathbf c}(\nu)^T \right),
\end{equation}
where ${\mathbf F}_N$ is an $N$-dimensional DFT matrix, $\odot$ is the Hadamard product, ${\mathbf X}_{\rm c} = [{\mathbf x}_{\rm c}(0),{\mathbf x}_{\rm c}(1),...,{\mathbf x}_{\rm c}(N_{s}-1)]$ with ${\mathbf x}_{\rm c}(n) = [{x}_{{\rm c},0}(n),{x}_{{\rm c},1}(n),...,{x}_{{\rm c},N_c-1}(p)]^T$, ${\mathbf b}(\tau') = [1, e^{-j2\pi \Delta f \tau'},...,e^{-j2\pi (N_c-1) \Delta f \tau'}]^T$ with $\tau'$ being the time delay, and ${\mathbf c}(\nu) = [1, e^{-j2\pi f_c T_{\rm c} \nu},...,e^{-j2\pi f_c (N_s-1) T_{\rm c} \nu}]^T$ with $f_c \nu$ being the Doppler shift. Then FFT could be applied before the detection of symbols $x_{{\rm c},m}(n)$ for $m=0,1,...,N_c-1$.

In most practical scenarios, the radio channel is both time- and frequency-dispersive, such that channel output spreads over time-frequency domains. Such channel distortion results in the so-called inter-symbol interference (ISI) and inter-channel interference (ICI) onto the received signals. By defining the time-frequency lattice based on symbol duration and subcarrier bandwidth, namely time-frequency plane, ISI and ICI can be reduced via well-localized 2D pulse shaping filters. Unfortunately, simultaneously sharply localizing a time- and frequency-limited signal on the time-frequency plane to well-concentrate its energy is impossible, as stated by Heisenberg Uncertainty Principle.

\subsection{Interplay between R\&C}
\subsubsection{Pulse Shaping for Radar and Communications}
Pulse shaping is essential for both R\&C to shape the waveform of the transmitted signal. Although signaling pulses serve a similar purpose in both cases, there are some key differences in their design and implementation. In communication systems, pulse-shaping is used primarily to minimize ISI and to control the bandwidth of the transmitted signal. This helps optimize the data rate, signal quality, and spectral efficiency. In radar systems, in addition to bandwidth control, pulse-shaping is also applied to control the sidelobes of the transmitted waveform. This helps to improve the range resolution and target detection capability. Furthermore, in communication systems, common pulse-shaping filters include the raised cosine filter, root raised cosine filter, Gaussian filter, and various others. These filters are chosen based on the specific modulation scheme, channel conditions, and system requirements. In radar systems, common pulse-shaping filters include the Hamming window, Blackman window, Chebyshev window, and Taylor window, among others. These filters are chosen based on the radar's specific requirements, such as the desired peak sidelobe level and range resolution.
\subsubsection{OFDM-based Radar vs. Delay-Doppler Communications} 
Multi-carrier techniques have been extensively used over the last decade for wideband systems. Examples include the SFW for radars and the OFDM for communications. It is worth noting that the OFDM signal can also be used for radar sensing, which is known as the communication-centric ISAC waveform that will be elaborated later. In such a system, the ISAC Tx transmits signals jointly for radar sensing and communicating with other communication systems by using the same OFDM signal, where each symbol is individually modulated with data belonging to a constellation. Accordingly, the OFDM blocks are individually processed at the Rx of the ISAC system. While the communication processing consists of extracting modulated data from each block, the radar processing consists of estimating the range-Doppler profile through the 2D-FFT operation \cite{sturm2011waveform}. As discussed in the previous subsection, the ISI and ICI cannot be fully eliminated in OFDM systems. To ease such issues, the recently developed orthogonal time-frequency space (OTFS) modulation proposed to use the delay-Doppler (DD)-based signal representation to convert the time-frequency channel responses into simple 2D time-invariant channel response \cite{hadani2018otfs}, thus alleviating the time-frequency selective effects. In such a case, the available signal propagation paths become physically explainable, observable, and probably predictable by, for example, moving object tracking strategies \cite{yuan2021integrated}. These key observations mandate the OTFS to be a novel ISAC SP paradigm which goes beyond separately performing R\&C SP on the DD and time-frequency domains.

\section{Scaling up the Antenna Array: The Road from Single Antenna to Massive MIMO}
In the last decade, the evolution of both R\&C systems has gained considerable spatial efficiency by scaling up the antenna arrays. The more antennas equipped at Tx/Rx, the more degrees of freedom (DoFs) that signaling strategies can be exploited from the propagation channel, and the better reliability can be achieved in the transmission. In this section, we investigate the evolution path of the array structure.
\subsection{Array Structure Evolution and Signal Models}
In general, an antenna array can be described by its response, a.k.a. steering vector, which is a vector function of angle parameters $\theta$, denoted as $\mathbf{a}\left(\theta\right)$. For an $N$-antenna uniform linear array (ULA) with antenna spacing $d$ and wavelength $\lambda$, the steering vector is expressed as
\begin{equation}\label{steering_vector}
{\mathbf{a}}\left( \theta  \right) = \left[ {1,{e^{ - j2\pi \frac{d}{\lambda }\sin \left( \theta  \right)}},{e^{ - j4\pi \frac{d}{\lambda }\sin \left( \theta  \right)}}, \ldots ,{e^{ - j\left( {N - 1} \right)\pi \frac{d}{\lambda }\sin \left( \theta  \right)}}} \right],
\end{equation}
where $\theta \in \left[-\pi, \pi\right]$, and $d$ is typically set as $\lambda/2$. Suppose that the radar or communication system is equipped with $N_t$ and $N_r$ antennas at its Tx and Rx, and that the signal arrives from $L$ resolvable paths. The general channel matrix for both R\&C can be modeled as
\begin{equation}\label{channel_matrix}
{\mathbf{H}} = \sum\nolimits_{l = 1}^L {{\alpha _l}{\mathbf{b}}\left( {{\theta _l}} \right){{\mathbf{a}}^T}\left( {{\phi _l}} \right)} 
\end{equation}
where $\alpha_l$, $\phi_l$, and $\theta_l$ are the channel coefficient, direction of departure (DoD), and direction of arrival (DoA) for the $l$th signal path, $\mathbf{a}\left(\phi\right) \in \mathbb{C}^{N_t \times 1}$ and $\mathbf{b}\left(\theta\right) \in \mathbb{C}^{N_r \times 1}$ are Tx and Rx steering vectors, respectively. The channel model (\ref{channel_matrix}) may represent $L$ resolvable point targets for radar, or $L$ propagation paths for communication. In the communication case, $\alpha_l$ is contributed by both the path-loss and small-scale fading effect. In the radar case, $\alpha_l$ may also be contributed by the radar cross-section (RCS) of the targets in addition to the round-trip path-loss, which follows the Swerling's target models \cite{swerling1960probability}.

\begin{figure*}
    \centering
    \includegraphics[width=2\columnwidth]{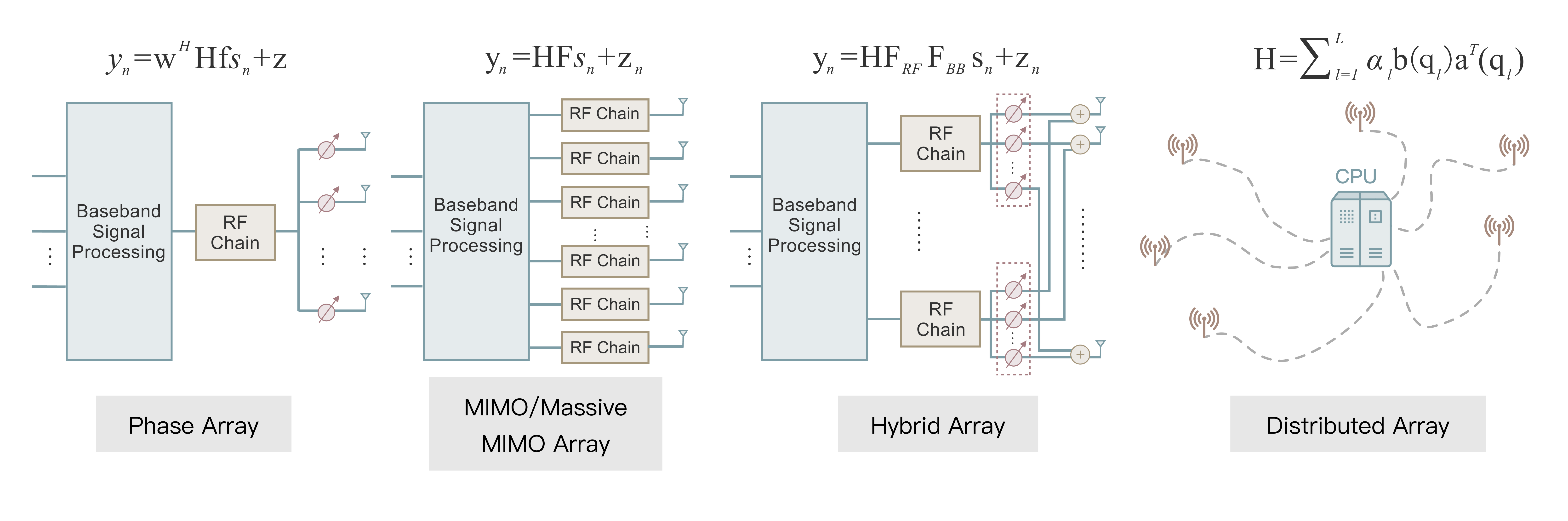}
    \caption{Antenna array evolution and signal models.}
    \label{fig: antenna_array}   
\end{figure*}

\textbf{Phased Array:} Having the capability of generating highly directive beam through rapid electronic phase control, phased-array techniques triggered various R\&C innovations. The phased array system, in its simplest form, consists of a single RF chain connected with multiple antennas through phase shifters. In other words, the signal transmitted over each antenna is a phase-shifted counterpart of the signal generated in the RF chain. If both the Tx and Rx are equipped with phased arrays, the discrete receive signal at time instant $n$ can be expressed as
\begin{equation}\label{phased_array}
{y_n} = {{\mathbf{w}}^H}\mathbf{H} {\mathbf{f}}{s_n} + {z_n},\;\forall n,
\end{equation}
where $s_n$ is the signal transmitted within the Tx's RF chain, ${\mathbf{f}} \in \mathbb{C}^{N_t \times 1}$ and ${\mathbf{w}} \in \mathbb{C}^{N_r \times 1}$ consist of the phase-shifters at the Tx and Rx, with each of their entries being constant-modulus, which are also known as the transmit beamformer and receive combiner, and are referred to as RF/analog beamforming.

\textbf{MIMO (Digital) Array:} In contrast to the phased array, the MIMO system is equipped with multiple RF chains, where each RF chain is connected to a single antenna port. The receive signal for a MIMO system can be modeled as
\begin{equation}\label{MIMO_array}
{{\mathbf{y}}_n} = \mathbf{H} {\mathbf{F}}{{\mathbf{s}}_n} + {{\mathbf{z}}_n},\;\forall n,
\end{equation}
where $\mathbf{s}_n \in \mathbb{C}^{K \times 1}$ and $\mathbf{y}_n \in \mathbb{C}^{N_r \times 1}$ are transmit and receive signal vectors at the Tx and Rx, respectively, with $K$ being the number of independent signals, and $\mathbf{F} \in \mathbb{C}^{N_t \times K}$ a digital precoder. In MIMO radar applications, $\mathbf{s}_n, \forall n$ are spatially orthogonal waveforms, and $\mathbf{F}$ may be designed to steer the signals to multiple directions simultaneously, or to keep the orthogonality for omni-directional searching. In MIMO communication applications, $\mathbf{F}$ may be designed to equalize or exploit the multi-path effect using various precoding techniques, e.g., zero-forcing (ZF) and MF precoding. MIMO communication technology was first patented in 1994 \cite{paulraj1994increasing}, which inspired the invention of the MIMO radar concept in 2003 \cite{bliss2003multiple}.

\textbf{Massive MIMO (mMIMO) Array:} When the antenna number grows extremely large, e.g., above 100, the MIMO system becomes an mMIMO system, or a large-scale antenna system. In this case, the steering vectors are asymptotically orthogonal to each other.
Moreover, in a richly scattering environment with large $L$, for ${N_t} \to \infty, N_t \gg N_r$, we have ${{\operatorname{var} \left( {{{\left\| {{{\mathbf{h}}_k}} \right\|}^2}} \right)} \mathord{\left/
 {\vphantom {{\operatorname{var} \left( {{{\left\| {{{\mathbf{h}}_k}} \right\|}^2}} \right)} {\mathbb{E}\left( {{{\left\| {{{\mathbf{h}}_k}} \right\|}^2}} \right)}}} \right.
 \kern-\nulldelimiterspace} {\mathbb{E}\left( {{{\left\| {{{\mathbf{h}}_k}} \right\|}^2}} \right)}} \to 0,\forall i$, and $\frac{1}{{{N_t}}}{\mathbf{H}}{{\mathbf{H}}^H} \approx {\mathbf{I}}_{N_r}$, which are known as the {\it {channel hardening}} and {\it{favorable propagation}} effects. While the basic signal model for mMIMO remains the same to (\ref{MIMO_array}), it has additional superiorities over small-scale MIMO \cite{5595728}. First, one may attain even more DoFs if equipping both Tx and Rx with mMIMO arrays. More importantly, the channel hardening effect improves the communication reliability by generating a nearly deterministic channel, which considerably simplifies the SP. Recent research has also shown that the mMIMO radar is able to detect a target via a single snapshot in the presence of disturbance with unknown statistics \cite{8962251}.

\textbf{Hybrid Array:} Massive MIMO achieves dramatic gains at the price of growing number of antennas and RF chains, incurring larger hardware costs. To that end, the hybrid analog-digital array was proposed as a promising solution \cite{6717211}. The hybrid array can be veiwed as a tradeoff between the phased-array and fully-digital MIMO array, as it connects fewer RF chains with massive antennas through phase-shifters or switches. Consider a hybrid array with $N_{RF}$ RF chains and $N_t$ antennas. The phase-shifter based design has the following signal model
\begin{equation}\label{HBF}
{{\mathbf{y}}_n} = {\mathbf{H}}{{\mathbf{F}}_{RF}}{{\mathbf{F}}_{BB}}{{\mathbf{s}}_n} + {{\mathbf{z}}_n},\;\forall n,
\end{equation}
where ${{\mathbf{F}}_{RF}} \in {\mathbb{C}^{{N_t} \times {N_{RF}}}}$ is the analog beamforming matrix containing constant-modulus entries representing phase-shifters, and ${{\mathbf{F}}_{BB}} \in {\mathbb{C}^{{N_{RF}} \times {K}}}$ is a digital precoder multiplexing $K$ data streams. The hybrid array is also known as the phased-MIMO structure in the radar community \cite{5419124}. In addition to reducing the cost for implementing the MIMO radar, it achieves a balance between phased-array and MIMO radars via harvesting performance gains from both. By partitioning the antenna array into different sub-arrays, phased-MIMO radar may formulate highly directional beams towards targets at each sub-array, improving the SNR of the echoes. In the meantime, it may also transmit orthogonal waveforms over different sub-arrays, thus reap the gain of waveform diversity.

\textbf{Distributed Array:} The continually growing demands for connectivity, coverage, and high-resolution sensing necessitate research of the distributed antenna array system for both R\&C. Instead of colocating the antennas in a compact space, distributed antennas are spread over a large area while connecting to a central processing unit (CPU), providing much higher probability of coverage and improved diversity gain. Distributed antenna systems have been extensively studied from the communication viewpoint under different names, including networked MIMO, coordinated multi-point (CoMP) and cell-free mMIMO \cite{7827017}. Its radar counterparts, on the other hand, are known as multi-static radar, and MIMO radar with widely separated antennas \cite{4408448}. The distributed array may also be described by its response, which, however, is no longer a function of the angle, rather it is a function of the coordinates of the targets or scatterers in each signal path. By denoting the coordinates of the $l$th target/scatterer as $\mathbf{q}_l = \left(x_l,y_l\right)$, the distributed channel matrix can be expressed as
\begin{equation}\label{distributed_channel}
    {\mathbf{H}} = \sum\nolimits_{l = 1}^L {{\alpha _l}{\mathbf{b}}\left( {{{\mathbf{q}}_l}} \right){{\mathbf{a}}^T}\left( {{{\mathbf{q}}_l}} \right)}.
\end{equation}
Note that the specific array geometry relies upon the overall deployment of the distributed system.

\subsection{Signal Processing for MIMO Radar and Communications}
\subsubsection{Colocated MIMO Radar}
With colocated antennas, MIMO radars can mimic beamformers utilizing low probability of intercept (LPI) waveforms. Rather than focusing energy on a target, the transmitted energy is evenly distributed in space \cite{4350230,4408448}. Compared to conventional phased-array beamforming, the loss of processing gain due to the uniform illumination is compensated by the gain in time, since there is no need to scan a narrow beam \cite{4350230,4408448}. The beamforming of classic colocated MIMO computes the correlations between the observation vectors from the previous step and the steering vectors corresponding to each azimuth/elevation on the grid defined by the array aperture. Then the targets can be detected in the angular domain. It is worth noting that a heuristic detection process, in which knowledge of the number of targets, clutter location, and so on, may help in discovering targets' positions \cite{zheng2019radar}. For example, if we know there are $M$ targets, then we can choose the $M$-strongest points in the targets profile. Alternatively, constant false alarm rate (CFAR) detectors determine a power threshold, above which a peak is considered to originate from a target so that a required false-alarm probability is achieved. 


\subsubsection{Distributed MIMO Radar} 
Widely separated transmit/receive antennas capture the spatial diversity of the target's RCS \cite{4408448}. Practical realization of phase coherency may be difficult, thus often necessitating non-coherent combining to perform target detection using the distributed apertures \cite{4408448}. It is shown that with non-coherent processing, a target's RCS spatial variations can be exploited to obtain a diversity gain for target detection and for estimation of various parameters, such as the DoA and Doppler. Again, the Swerling models \cite{swerling1960probability} can be used to represent the statistical RCS fluctuations as a function of the target decorrelation time. In distributed MIMO radars, a multi-dimensional signal space is created when the returns from multiple scatterers or targets combine to generate a rich backscatter. By exploiting the spatial dimension, MIMO radar with widely separated antennas may overcome bandwidth limitations and support high resolution target localization \cite{4408448}.



\subsubsection{MIMO Communications}
MIMO communication has been playing a critical role in the cellular and Wi-Fi systems since 2010s, the beginning of the 4G era. Early signal processing methods focused on the single-user MIMO (SU-MIMO) communication, where a multi-antenna base station (BS) serves a single- or multi-antenna user, which is also known as the point-to-point (P2P) MIMO channel. 
In addition to multiplexing more data streams, MIMO array is also able to serve multiple users over the same time-frequency resource block, which is known as multi-user MIMO (MU-MIMO) communications technology. Comparing to SU-MIMO, MU-MIMO configuration offers significant complexity reduction at the users' side. 
For MU-MIMO systems, the coordinated signal detection at the receivers' side is not as straightforward as in the SU-MIMO, since cooperation among users is difficult. Therefore, the BS needs to pre-cancel the interference by employing various precoding methods, which also simplifies the SP at users' side. While dirty paper coding (DPC) is capacity-achieving, it suffers from high complexity \cite{cho2010mimo}. Therefore, sub-optimal linear precoders are more commonly employed in practical systems. 

\subsection{Interplay between R\&C}
\subsubsection{Multiplexing vs. Diversity}
The expansion of the antenna array brings diversity and multiplexing gains, which are cornerstones of MIMO communication theory. Transmit or receive diversity is a means to combat deep fading by creating different propagation paths through the Tx-Rx antenna pairs. Multiplexing, on the other hand, {\it{exploits}} the DoFs provided by multi-path propagation environment through sending different data streams over independent subchannels. In 2003, L. Zheng and D. Tse revealed that there is an inherent tradeoff between the two gains, namely, diversity-multiplexing tradeoff (DMT) \cite{1197843}. For an i.i.d. Rayleigh MIMO channel $\mathbf{H}_{\rm c} \in \mathbb{C}^{N_r \times N_t}$, the maximum diversity gain and multiplexing gain are $N_tN_r$ and $\min\left\{N_t, N_r\right\}$, respectively. From a broader viewpoint, DMT is essentially a tradeoff between reliability and efficiency. 

The spirit of MIMO radar signal processing can be interpreted in a similar manner. On one hand, colocated MIMO radar possesses the superior attribute of {\it{waveform diversity}}, which means that diverse waveforms are flexibly emitted through different antennas. Waveform diversity may be implemented in either baseband or RF band, e.g., through phase coding or frequency coding. It significantly improves the parameter identifiability compared to its phased-array counterpart. That is, the colocated MIMO radar is able to uniquely identify up to $\mathcal{O}\left(N_tN_r\right)$ targets, which is $N_t$ times of that of the phased-array radar \cite{4350230}. This connects more closely to the multiplexing gain in communications. On the other hand, distributed MIMO radar provides {\it{target RCS diversity}}. By widely spreading the antennas, distributed MIMO radar is able to observe a target from different directions, thus to provide a stable sensing performance by overcoming the drastic RCS fluctuations in high-mobility targets \cite{4408448}. 

The above discussion reflects again the signals-and-systems duality. Since the signals and systems are interchangeable, we may view radar target channels as ``signals'', and radar waveforms as ``systems''. While the basic model for MIMO communications is that multiple data streams (signals) are transmitted through multiple spatial channels (systems), the model for MIMO radar is, conversely, that multiple target channels (signals) pass through diverse waveforms (systems). This duality creates the interesting interplay between R\&C, and may imply more essential connections and tradeoffs in ISAC systems.

\subsubsection{Statistical vs. Geometrical Channel Representations}
Most of the MIMO radar channels are geometrically modeled, as the ultimate goal of the radar is to extract the physical parameters of targets. The MIMO communication channel, on the other hand, can be modeled either statistically or geometrically, depending on the specific scenarios and systems. The distinct models of the same channel are representations in different {\it{coordinate systems}}. For instance, an $N_r \times N_t$ communication channel matrix ${{\mathbf{H}}_{\rm c}}$ may be generally seen as a point in the Euclidean space $\mathbb{C}^{N_r \times N_t}$. If it is geometrically modeled, then it may be viewed as a point in a subspace spanned by steering vectors $\mathbf{a}\left(\phi_l\right)$ and $\mathbf{b}\left(\theta_l\right)$. In sub-6 GHz bands with richly scattering environments, the small-scale MIMO channel is modeled as an unstructured matrix subject to certain distributions, e.g., Rayleigh and Rician distributions, since the number of propagation paths could be far greater than that of the channel entries. In such a case, the communication channel estimation task is to recover all the entries in ${{\mathbf{H}}_{\rm c}}$. In mmWave and THz bands with much fewer propagation paths than antennas, the massive MIMO channel is well characterized by geometrical clustered model such as the Saleh-Valenzuela model, which enables beamspace signal processing for mmWave and THz communications that mimics MIMO radar signal processing. In fact, beam training and tracking in mmWave and THz communications may be analogously viewed as target searching and tracking, all of which can be operated on a hybrid array based RF platform. This also builds a solid foundation to merge R\&C into a single system by ISAC technologies.


\begin{figure}
    \centering
    \includegraphics[width=\columnwidth]{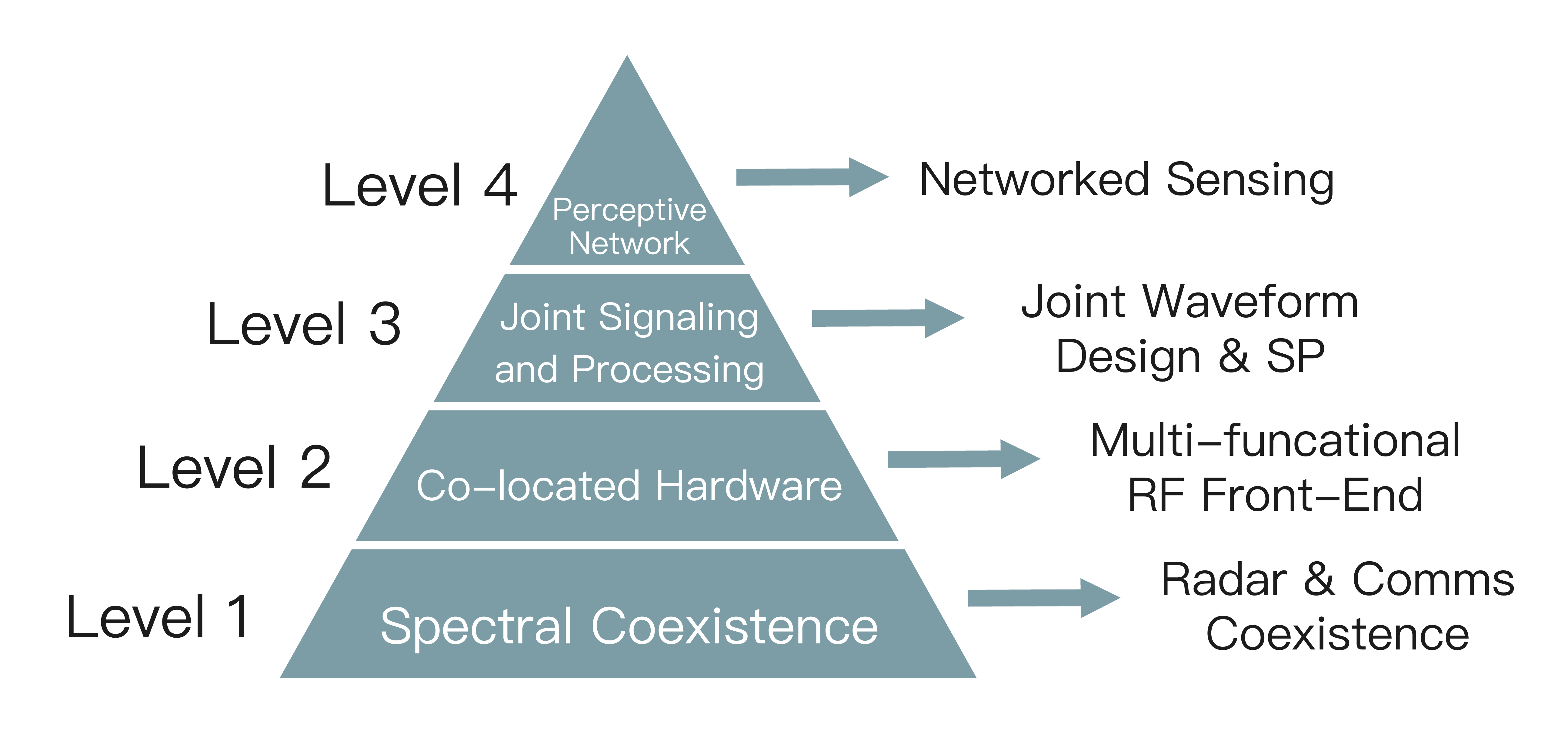}
    \caption{Evolution path for ISAC technologies.}
    \label{fig: ISAC_path}
\end{figure}
\section{Integrated Sensing and Communications: The Road from Separation to Integration}
\subsection{ISAC: From Competitve Coexistence to Co-design}
The ubiquitous deployment of R\&C systems leads to severe competition over various resource domains. To date, both technologies exhibit explosively growing demands for spectral and spatial resources, and are thus evolving towards higher frequencies and larger antenna arrays.  As exemplified in Sec. III-A, a variety of R\&C systems have to cohibitate within multiple frequency bands, which, inevitably, incurs significant mutual interference between the two functionalities \cite{zheng2019radar,7953658}. To ensure harmonious coexistence between R\&C, orthogonal resource allocation became a viable approach. Nevertheless, orthogonal allocation results in low resource efficiency for both R\&C. Aiming for fully excavating the potential of the limited wireless resources, e.g., bandwidth, and to enable the co-design of the R\&C functionalities, ISAC was proposed as a key technology for both next-generation wireless networks and radar systems.

The technological vision of ISAC can be divided into four levels, as shown in Fig. {\ref{fig: ISAC_path}}. The first level is to share the spectral resources between individual R\&C systems without interfering with each other. At the second level, the R\&C functionalities may be deployed on the same hardware platform. At the third level, wireless resources may be fully reused between R\&C via a common waveform, a single transmitting device, and a unified signal processing framework. Finally, at the fourth level, both R\&C can share a common networking infrastructure, constructing a {\it{perceptive network}} to serve for both sensing and communications functionalities. This underpins a large number of emerging IoT, 5G-Advanced and 6G applications that require high-quality communication, sensing, and localization services \cite{9737357}. 


During the past three decades, the development of ISAC has been supported by a number of governmental projects worldwide, among which the most influential ones are the ``Advanced Multifunction Radio Frequency Concept (AMRFC)'' program initiated by the Office of Naval Research (ONR) of the US in the 1990s, and the  ``Shared Spectrum Access for Radar and Communications (SSPARC)'' project funded by the Defense Advanced Research Projects Agency (DARPA) of the US in the 2010s \cite{8999605}. While both projects were motivated by the need of sharing resources between R\&C, the AMRFC mainly focused on colocating multi-functional modules (radar, communications, and electronic warfare) on the same RF front-ends, and the SSPARC aimed for releasing part of the sub-6 GHz spectrum from the radar bands for shared use between R\&C. Most of the technical outcome of these projects formulating the Level-1 to Level-3 ISAC approaches. In the 2020s, networked sensing (Level-4 ISAC) was recognized by major enterprises in the communications industry (Huawei, Ericsson, ZTE, Intel, and Nokia) as one of the core air interface technologies for WiFi-7, 5G-A and 6G \cite{9737357}. In 2020, IEEE 802.11 formed the 802.11bf task group to realize WLAN sensing in WiFi-7, which is expected to be commercialized in 2024 \cite{9941042}. In 2022, 3GPP established the first study item on ISAC towards Rel-19 standards for 5G-A \cite{3gpp.22.837}.

To fully realize the promise of the ISAC technology, advanced SP techniques are indispensable. In this section, we briefly review the recent research progress on the SP for ISAC. In particular, we focus on the Level-3 and 4 where a unified signaling strategy is designed to serve for the dual purposes of R\&C.

\subsection{ISAC Signal Processing}
We investigate the linear Gaussian models considered in Sec. II. The only difference is that a unified ISAC signal $\mathbf{S}$ is employed for both R\&C, leading to 
\begin{equation}\label{ISAC_model1}
\begin{gathered}
  \text{Radar Signal Model: } {{\mathbf{Y}}_{\rm r}} = {{\mathbf{H}}_{\rm r}}\left( {\bm{\eta }} \right){{\mathbf{S}}} + {{\mathbf{Z}}_{\rm r}}; \hfill \\ \text{Comms Signal Model: } {{\mathbf{Y}}_{\rm c}} = {{\mathbf{H}}_{\rm c}}{{\mathbf{S}}} + {{\mathbf{Z}}_{\rm c}}, \hfill \\
\end{gathered} 
\end{equation}
where $\mathbf{S}$ is a discrete representation of the ISAC signal. We highlight that (\ref{ISAC_model1}) are abstractions for many existing ISAC models. That is, an ISAC Tx transmits a signal $\mathbf{S}$ to communicate information while detecting targets. For radar sensing applications, the radar Rx observes $\mathbf{Y}_{\rm r}$, and wishes to extract an estimate of $\bm{\eta}$ with the knowledge of the reference waveform $\mathbf{S}$, which is known to both the ISAC Tx and radar Rx. For communication applications, on the other hand, the communication Rx observes $\mathbf{Y}_{\rm c}$, and wishes to recover $\mathbf{S}$, which is unknown to the communication Rx. 

\begin{figure*}
    \centering
    \includegraphics[width=2\columnwidth]{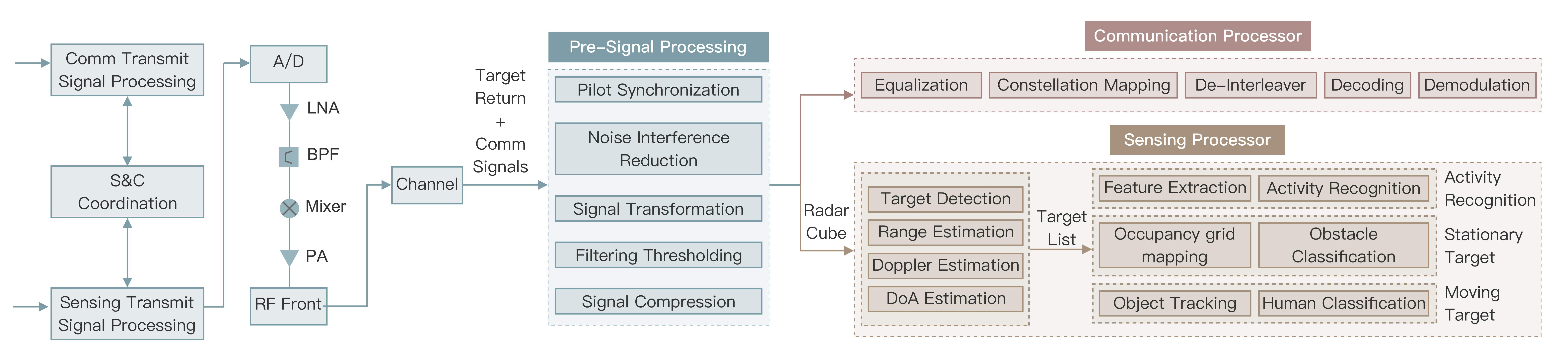}
    \caption{General signal processing framework for ISAC systems.}
    \label{fig: SP_framework}
\end{figure*}

\begin{figure}
    \centering
    \includegraphics[width=\columnwidth]{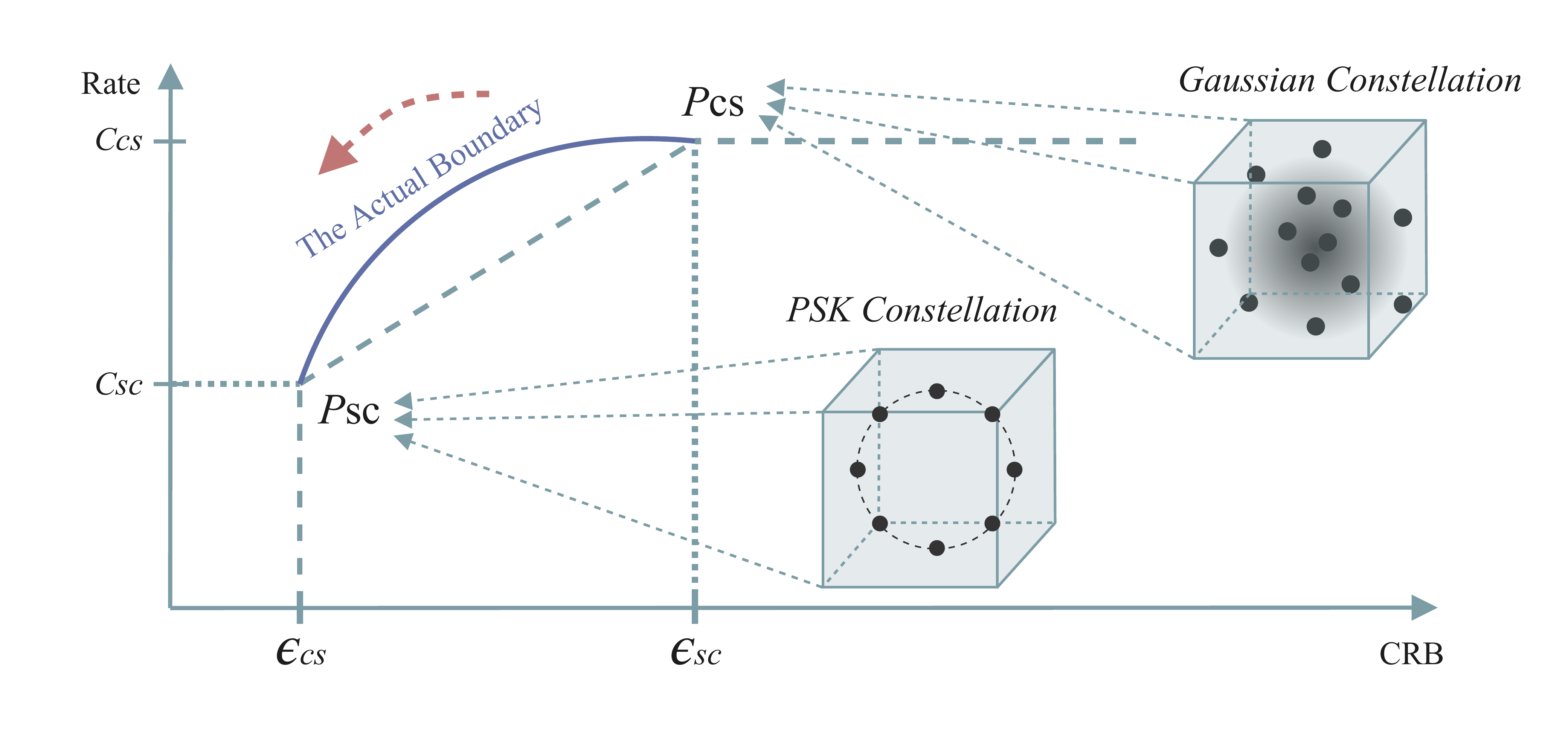}
    \caption{Performance tradeoff between R\&C.}
    \label{fig: RC Limits}
\end{figure}

A generic ISAC SP framework is shown in Fig. \ref{fig: SP_framework}, where the R\&C functionalities are jointly coordinated at the ISAC Tx to form a baseband ISAC signal. After being up-converted to the RF band, the signal propagates through the R\&C channels and arrives at the Rx. The received signal, which may consist of both target and communication information, first goes through a pre-processing procedure including synchronization, separation, filtering and transformation, and is then processed following the regular R\&C SP pipelines. ISAC SP is rather different from the individual R\&C SP. That is, when the wireless resources are shared between R\&C, there exists an intrinsic {\it performance tradeoff} as their design objectives are distinct or even contratictory to each other. As shown in Fig. {\ref{fig: RC Limits}}, such a tradeoff can be framed as the Pareto frontier in terms of different R\&C performance metrics, e.g., radar's CRB and communication rate. The complete characterization of such a Pareto frontier still remains wide open. The two corner points, $P_{CS}$ and $P_{SC}$, represent the communication-optimal and radar-optimal performance, with the corresponding achievable rate-CRB pairs denoted by $\left(C_{CS}, \epsilon_{CS}\right)$ and $\left(C_{SC}, \epsilon_{SC}\right)$, respectively. This results in three categories of ISAC SP designs, i.e., communication-centric, radar-centric, and joint design, which target on approaching the points $P_{CS}$ and $P_{SC}$, and the Pareto frontier in between, respectively.

\subsubsection{Communication-Centric Design}
Communication-centric design (CCD) simply implements the radar sensing functionality over an existing or even commercialized communication waveform, in which case the communication functionality has the priority. The most representative CCD approach is the OFDM-based ISAC signaling, which directly exploits the OFDM communication waveform to simultaneously accomplish R\&C tasks \cite{sturm2011waveform,9529026}. Assume that the ISAC Tx emits the OFDM signal to communicate with a user, while sensing a point target with delay $\tau$ and Doppler $\nu$. After receiving the echo signal reflected from the target, the radar Rx, which is colocated with the ISAC Tx, samples at each OFDM symbol, followed by a block-wise FFT processing. The resultant discrete signal can be arranged into a matrix, with its $\left(n,m\right)$-th entry associating with the $n$th symbol at the $m$th subcarrier, given as
\begin{equation}\label{OFDM_echo}
{y_{n,m}} = {\alpha _{n,m}}{x_{n,m}}{e^{ - j2\pi \left( {m - 1} \right)\Delta f\tau }}{e^{j2\pi {\nu}\left( {n - 1} \right){T_c}}} + {z_{n,m}},
\end{equation}
where $\alpha_{n,m}$ and $z_{n,m}$ are the channel coefficient and noise. The random communication data ${x_{n,m}}$ impose a negative impact on radar sensing, which can be simply mitigated by element-wise division
\begin{equation}\label{element_wise_division}
\begin{gathered}
  {{\tilde y}_{n,m}} = {{{y_{n,m}}} \mathord{\left/
 {\vphantom {{{y_{n,m}}} {{x_{n,m}}}}} \right.
 \kern-\nulldelimiterspace} {{x_{n,m}}}}\hfill \\ \quad\quad\;= {\alpha _{n,m}}{e^{ - j2\pi \left( {m - 1} \right)\Delta f\tau }}{e^{j2\pi {\nu}\left( {n - 1} \right){T_c}}} + {{{z_{n,m}}} \mathord{\left/
 {\vphantom {{{z_{n,m}}} {{x_{n,m}}}}} \right.
 \kern-\nulldelimiterspace} {{x_{n,m}}}}. 
\end{gathered}
\end{equation}
Then, a 2D-FFT can be applied to (\ref{element_wise_division}) to get the DD profile of the target.

\subsubsection{Radar-Centric Design}
In contrast to CCD schemes, radar-centric design (RCD) aims at implementing the communication capability over existing radar infrastructures, targeting on approaching the performance at $P_{SC}$. Since the classical radar waveform contains no information, RCD schemes are also referred to as information embedding approaches in the literature, namely, the communication data is embedded into the radar waveform, in a way that will not unduly degrade the sensing performance. Early RCD schemes have mainly focused on exploiting the LFM signal as an information carrier \cite{saddik2007ultra}. In addition to the conventional modulation formats including amplitude, phase, and frequency shift keyings, LFM signals have another design DoF, i.e., the slope that the frequency increases with the time, which may also be utilized for data embedding. To fully guarantee the radar performance, recent research proposed to realize ISAC by {\it{index modulation (IM)}}, which was first proposed in \cite{7485316} for MIMO radar transmiting orthogonal waveforms. In such a case, the communication information was conveyed by shuffling the waveforms across multiple antennas, which does not break the orthogonality. As a step forward, more recent RCD schemes implement IM-based ISAC signaling on the carrier-agile phased arrary radar (CAESAR), namely, the multi-carrier agile joint radar-communication (MAJoRCom) system \cite{9093221}. During each PRI, the MAJoRCom randomly selects the carrier frequencies from a frequency set, and randomly allocates these frequencies to each antenna, which again keeps the orthgonality unaffected.

\subsubsection{Joint Design}
As discussed above, CCD and RCD schemes attempt to approach the performance of $P_{CS}$ and $P_{SC}$, which may be implemented in existing communication and radar systems, respectively. However, they lack the flexibility to formulate a scalable tradeoff between R\&C, or, equivalently, to approach the performance of an arbitray point on the Pareto frontier in Fig. \ref{fig: RC Limits}. To resolve this issue, joint design (JD) based ISAC signaling becomes a promising strategy, which are often conceived through convex optimization techniques \cite{liu2021CRB}. Consider a MIMO ISAC BS that serves $K_u$ single-antenna users while detecting a point target locating at an angle $\theta$. A ISAC signal $\mathbf{S}$ constrained by the energy $E_T$ can be obtained by solving the below angle CRB minimization problem under the sum-rate constraint
\begin{equation}\label{ISAC_precoding}
  \mathop {\min }\limits_{\mathbf{S}} \;\operatorname{CRB} \left( \theta  \right) \;\;\text{s.t.}\;\;\mathop \sum \nolimits_{k = 1}^{K_u}R_k \ge R_0,\forall k,\;\left\| {\mathbf{S}} \right\|_F^2 \le {E_T}, \hfill \\ 
\end{equation}
where $R_k$ is the achievable rate for the $k$th user, and $R_0$ is a pre-defined sum-rate threshold. The Pareto frontier between R\&C can be obtained by increasing $R_0$, which leads to increased objective CRB.

\subsection{Interplay between R\&C}
From the above ISAC SP strategies, it is interesting to note that there is a two-fold tradeoff between R\&C, namely, {\it{deterministic vs. random tradeoff (DRT)}} and {\it{subspace tradeoff (ST)}}.
\subsubsection{Deterministic vs. Random Tradeoff}
Communication systems require random signals to convey as much information as possible, whereas radar systems prefer deterministic signals for achieving stable sensing performance. This has been an intuitive insight consistent with both the engineers' experience and R\&C SP theory. For instance, constellation shaping for communications always target on approximating a Gaussian distribution, thus to approach the Shannon capacity. Radar systems, on the other hand, prefer to transmit constant-modulus waveforms at the maximum available power budget, which motivates the use of phase-coded signals. For clarity, this concept has been shown in Fig. \ref{fig: RC Limits}.

The DRT has also been reflected in the above CCD and RCD approaches. For OFDM-based CCD signaling, the element-wise division of the random data changes the statistical characteristics of the noise across the symbols and subcarriers, imposing performance loss to the thresholding and peak detection in the 2D-FFT processing. To tackle this issue, a natural idea is to transmit PSK modulated data, which only rotates the phase of the circularly symmetric Gaussian noise without changing its distribution. For IM-based RCD scheme, the radar transmits communication data by the random selections of waveforms across the antennas, i.e., the information is carried by permutation or selection matrices, while keeping the radar waveform orthogonality unchanged. In both cases, the communication rate can be increased by embedding more random data (exploiting more DoFs) into the ISAC signal, which is however at the price of deteriorated radar sensing performance.

\begin{figure*}
    \centering
    \includegraphics[width=1.7\columnwidth]{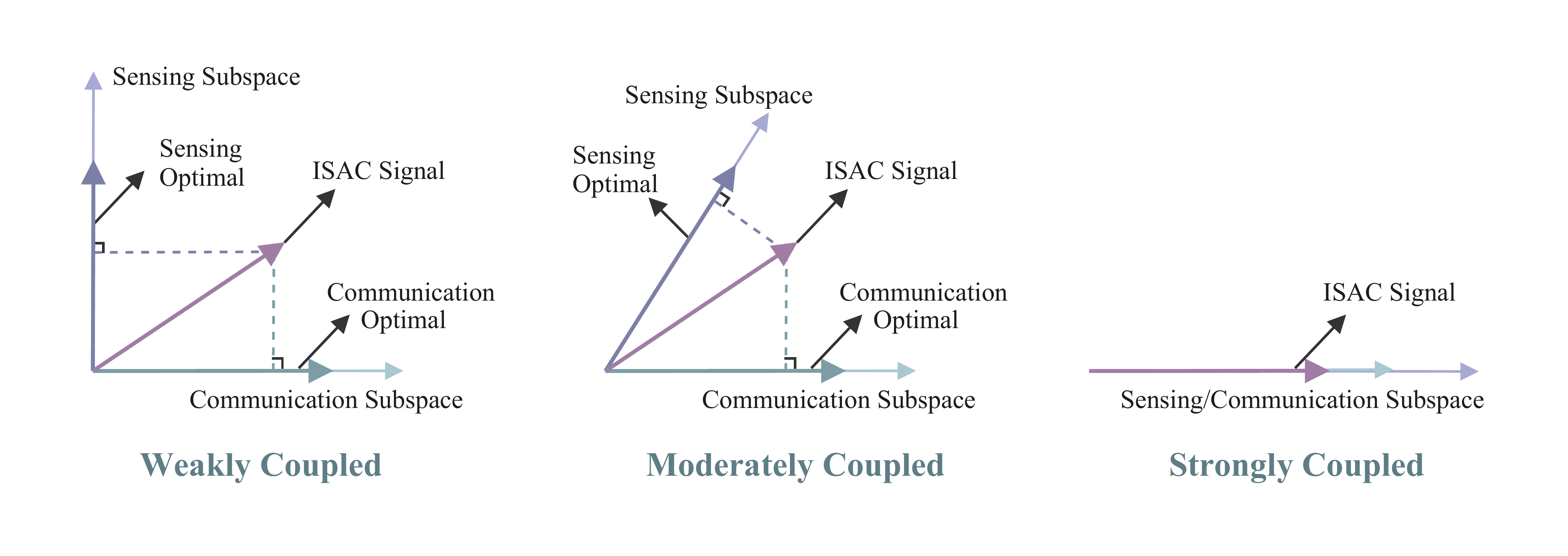}
    \caption{Subspace tradeoff and coupling effect in ISAC systems.}
    \label{fig: RC subspace}
\end{figure*}

\subsubsection{Subspace Tradeoff}
Another fundamental tradeoff in ISAC is the subspace tradeoff. The column vectors of R\&C channel matrices $\mathbf{H}_{\rm{r}},\mathbf{H}_{\rm{c}}$ span the sensing and communication subspaces. In order to fully radiate the transmit power towards targets/users, radar-optimal and communication-optimal signals should align to the two subspaces, respectively. Consequently, the R\&C performance can be balanced in an ISAC system by allocating resources into the two subspaces. Apparently, if two subspaces are partially overlapped, then resources allocated to the intersection are shared between R\&C, improving the efficiency. On the contrary, if two subspaces are orthogonal to each other, no resources can be reused, leading to zero performance gain. Based on the overlapped degree of two subspaces, one may categorize R\&C channels as weakly coupled, moderately coupled, and strongly coupled scenarios, which are intuitively illustrated in Fig. \ref{fig: RC subspace}. The higher coupling degree between two subspaces results in better tradeoff performance, as more resources are reused between R\&C.

\begin{figure}
    \centering
    \includegraphics[width=0.8\columnwidth]{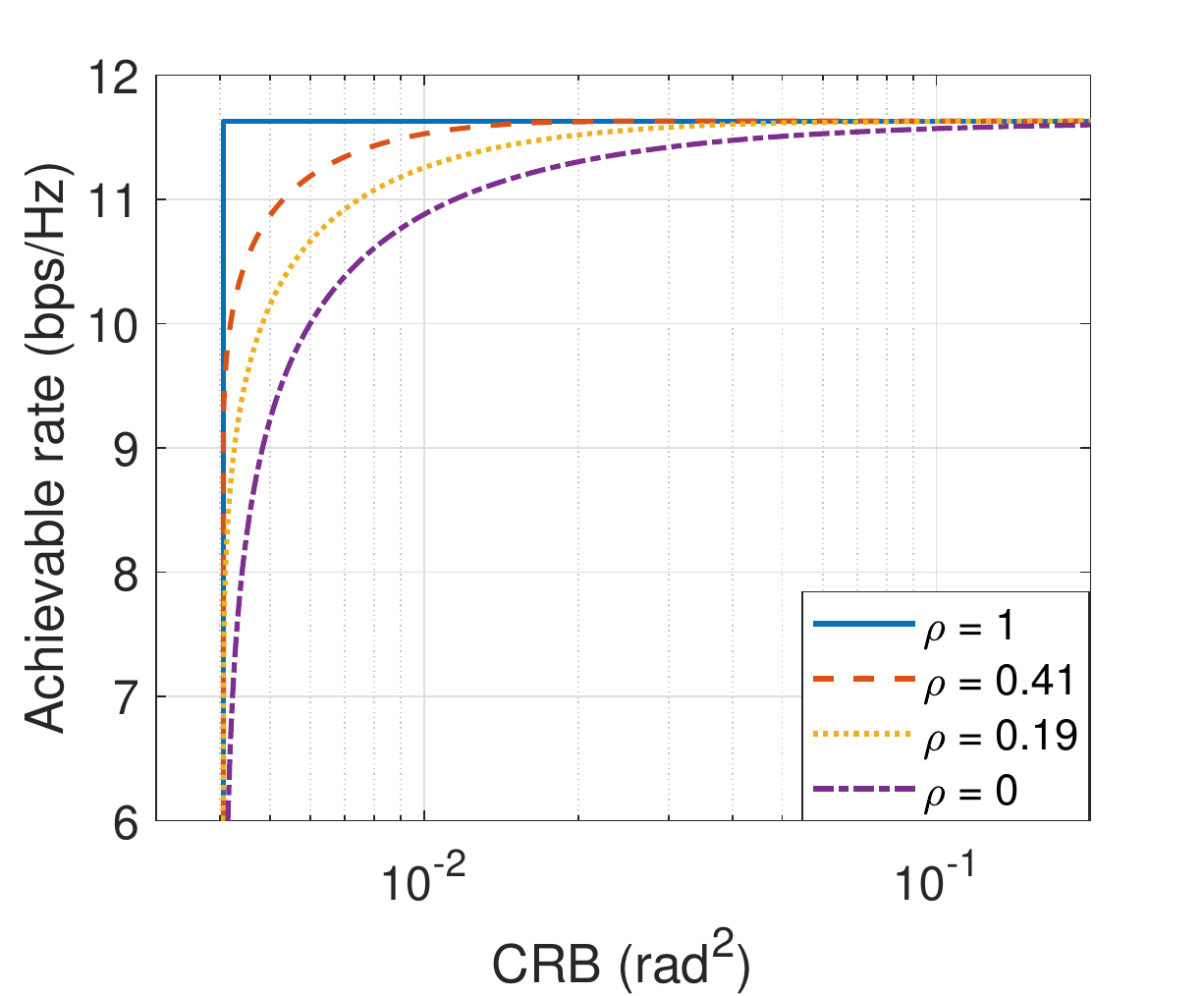}
    \caption{Radar and communication performance tradeoff under different correlation coefficients.}
    \label{fig: RC correlations}
\end{figure}

The ST can be observed in the JD signaling scheme discussed in (\ref{ISAC_precoding}). That is, by increasing the communication sum-rate threshold $R_0$, more signal power is transmitted towards the directions of communication users, while less power is radiated to sense the target, resulting in a higher CRB. To illustrate this, we show a numerical example of solving problem (\ref{ISAC_precoding}) in Fig. \ref{fig: RC correlations}, for a single-target single-user scenario. In particular, we consider the correlation coefficient between the communication channel $\mathbf{h}_c$ and the target steering vector $\mathbf{a}\left(\theta\right)$, defined as $\rho  = \frac{{\left| {{\mathbf{h}}_c^H{\mathbf{a}}\left( \theta  \right)} \right|}}{{\left\| {{{\mathbf{h}}_c}} \right\|\left\| {{\mathbf{a}}\left( \theta  \right)} \right\|}}$. By varying the SINR constraint of the user, we observe that the resultant ISAC signal indeed formulates a scalable tradeoff between the radar CRB and the communication achievable rate, where the ISAC signal rotates from the communication subspace to the sensing subspace. More interestingly, by increasing the correlation coefficient $\rho$ from 0 to 1, the ISAC tradeoff performance becomes better, which is consistent with our analysis on weakly, moderately, and strongly correlated subspaces. That is, higher correlation between two subspaces indicates that more resources can be shared between R\&C. In the extreme case of $\rho = 1$, both R\&C performance reaches to their optimum without jeopardizing one another. This is because the two subspaces are fully aligned to each other, and the signal resources can be fully reused between R\&C, leading to the maximum gain.

\section{Open Challenges and Future Research Directions}
Although ISAC has been well-investigated from various facets in recent years, there are still many open challenges that remain widely unexplored. Here we overview some of the open problems in fundamental tradeoff, signal processing, and networking aspects, where tremendous research efforts are needed.
\subsubsection{Full Characterization of the ISAC Performance Tradeoff}
Characterizing the ISAC performance tradeoff is a multi-objective functional optimization problem by its nature. Nevertheless, the current results are only able to depict the performance at the two corner points \cite{xiong2022flowing_1}. It is unclear where the exact Pareto frontier lies in Fig. \ref{fig: RC Limits}, and what are the optimal signaling strategies to achieve that boundary. Moreover, the research on fundamental ISAC tradeoff in more practical scenarios, e.g., multi-user multi-target regime, is still at its early stage, where tighter estimation-theoretical bounds and multi-user capacity region need to be jointly considered.
\subsubsection{Practical ISAC Signal Processing}
Most of the current ISAC signaling schemes were proposed under ideal assumptions. However, there are a large number of practical constraints that prevent the implementation of these ISAC designs. For instance, CCD approaches that adopt standardized communication waveform, e.g., 5G NR, face the challenges of insufficient bandwidth and high peak-to-average power ratio (PAPR), which leads to severe performance loss of radar sensing. In addition to that, the imperfection of hardware components, e.g., quantized phase shifters and uncalibrated antenna arrays, also need to be taken into account in designing practical ISAC SP pipelines.
\subsubsection{Networked ISAC}
Current state-of-the-art research mainly concentrates on the SP for single-node ISAC systems. To realize networked ISAC using the commercialized networking infrastructures, which are not originally tailored for radar sensing, a series of SP challenges need to be carefully coped with. For instance, clock-level network synchronization is needed to achieve high sensing accuracy. Moreover, in order to detect short-range targets, e.g., humans and vehicles, the future ISAC BS should operate in full-duplex mode to avoid self-interference between the transmit signal and target return. Equipping the network with ubiquitous sensing capabilities has also raised concerns on security and privacy issues, which needs to be carefully coped with in future ISAC systems.

\section{Conclusion}
In this paper, we overviewed the technological evolution of R\&C from an SP viewpoint. We first focused our discussion on the general principles and fundamental SP techniques for both R\&C. We then introduced two main trends and the resulting signal processing schemes in the historical development of R\&C, namely, the increase of frequencies and bandwidths, and the expansion of the antenna arrays. Following these two trends, we provided a detailed discussion on the recent progress of SP techniques for ISAC systems. Finally, we identified a number of major open challenges in ISAC technologies.

Although being two long-established disciplines, the story of R\&C will continue in the foreseeable future. In particular, ISAC, the marriage between R\&C, will have large impact on modern society. 

\section*{Acknowledgement}
This work is supported in part by the National Natural Science Foundation of China (62101234); and Guangdong Pearl River Talent Project (2021QN02X128). Le Zheng is the corresponding author.

	\bibliographystyle{IEEEtran}
	\bibliography{IEEEabrv,references_SPM,references,database}

\end{document}